\documentclass[preprint*,notoc]{JHEP3}

\usepackage{epsfig,multicol}
\newcommand\fverb{\setbox\pippobox=\hbox\bgroup\verb}
\newcommand\fverbdo{\egroup\medskip\noindent%
            \fbox{\unhbox\pippobox}\ }
\newcommand\fverbit{\egroup\item[\fbox{\unhbox\pippobox}]}
\newbox\pippobox

\title{Chiral Symmetry in Light-front QCD}
\author{Meng-Hsiu Wu\\
    Department of Physics, National Cheng Kung
        University, Tainan, 701, Taiwan\\
    E-mail: \email{mhwu@hepth.phys.ncku.edu.tw}}
\author{Wei-Min Zhang\\
    Department of Physics, National Cheng Kung
        University, Tainan, 701, Taiwan\\
    and National Center for Theoretical Science, Taiwan \\
    E-mail: \email{wzhang@mail.ncku.edu.tw}}

\preprint{\hepph{0310095}}    
\abstract{ The definition of chiral transformations in light-front
field theory is very different from the conventional form in
equal-time formalism. We study the consistency of chiral
transformations and chiral symmetry in light-front QCD and derive
a complete new light-front axial-vector current for QCD. The
breaking of chiral symmetry in light-front QCD is only associated
with helicity flip interaction between quarks and gluons.
Remarkably, the new axial-vector current does not contain the pion
pole part so that the associate chiral charge smoothly describes
pion transitions for various hadronic processes.}


\begin{document}
\def\d{{\rm d}}
\def\be{\begin{eqnarray}}
\def\ee{\end{eqnarray}}
\def\nn{\nonumber}
\section{Introduction}

In the investigations of the nonperturbative QCD in light-front
theory, the most difficult problem is how to provide a clear
picture of the dynamical chiral symmetry breaking and the quark
confinement with a trivial light-front QCD vacuum \cite{WWHZPG}.
The trivial vacuum in light-front theory is obtained by removing
the longitudinal zero modes which can be easily realized by
imposing an explicit cutoff on small longitudinal momentum
\cite{WWHZPG} or via a suitable prescription on the operator
$1/\partial^+$ \cite{ZH}. Both methods give a specific
regularization of the longitudinal momentum going to be zero. It
is then natural to ask how the spontaneous chiral symmetry
breaking (which is a nontrivial vacuum effect of QCD in common
understanding and plays a crucial role in the description of low
energy hadronic physics) can be manifested in light-front
formulation. A few years ago, Wilson pointed out that the concept
of spontaneous chiral symmetry breaking may be realized via an
explicit symmetry breaking in light-front theory induced from
counterterms of removing longitudinal infrared momentum divergence
\cite{Wilson}, but up to date a practical scheme of this procedure
has not been carried out yet.

In fact, chiral symmetry itself is not a good dynamical symmetry
for strong interaction. Mechanism of dynamical symmetry breaking
can manifest differently in different space-time frameworks
\cite{BE,MB}. Therefore, before we attempt to derive the
counterterms related to chiral symmetry breaking through a
reliable renormalization procedure for the light-front QCD
Hamiltonian, it is certainly helpful to understand precisely the
concept of chiral symmetry and its breaking in light-front theory.
In this paper, we shall study chiral transformations in
light-front theory and the corresponding divergence of
axial-vector current for light-front QCD. We will then further
explore its relation with pionic dynamics and the associated
physical implication in hadronic structure.

If we do not take into account the anomaly, the axial-vector
current, defined by $j_5^{\mu}\equiv\bar{\psi}\gamma^{\mu}\gamma_5\psi$,
satisfies the following equation:
\be
\partial_{\mu}j_5^{\mu}=2im\bar{\psi}\gamma_5\psi,
\label{DIVEQ}
\ee
namely, the axial-vector current is not
conserved for massive fermion theory no matter the fundamental
theory is a free theory or an interacting theory. The divergence
of axial-vector current is only related to fermion mass and is
independent on the detailed interacting structure of the
underlying particles. However, in light-front theory, a fermion
field is separated into an up and a down components,
$\psi=\psi_++\psi_-$, and only the up component field ($\psi_+$)
is a dynamical variable. Thus, chiral dynamics of a fermion in
light-front theory might behave differently from that in
equal-time formulation. In fact, it was noted long time ago
\cite{CHKT} that since the fermion field $\psi$ mix the left- and
right-handed components in light-front theory, the usual
chiral transformation can only be applied to the up component
$\psi_+$. The transformation for down component $\psi_-$ is
determined by the light-front fermion constraint. As a result, it
can be shown that the standard axial-vector current $j_5^{\mu}$ is
not a Noether current associated with such a light-front chiral
transformation \cite{DM}. The new light-front axial-vector current
with respect to the light-front chiral transformation obeys a
divergence equation that differs from eq.~(\ref{DIVEQ}). In recent
years, similar idea has been applied to chiral Yukawa model,
Nambu-Jona-Lasinio model \cite{IM} and light-front O(2) sigma
model with fermions \cite{MV2001}, but up to date none has
explored seriously about chiral dynamics for light-front QCD from
this point of view. Using such a consistent light-front chiral
transformation, we find that the conservation law of the
axial-vector current in light-front theory is broken only by the
helicity flip interaction alone in the QCD Lagrangian and the new
light-front axial-vector current does not contain the pion-pole
contribution. This may provide us a new look about the chiral
symmetry and chiral dynamics of QCD in light-front theory.

The paper is organized as follows: In Section \ref{ctilfqcd}, we
explore in details chiral transformations in light-front
theory and chiral symmetry in light-front QCD. In Section
\ref{ca}, we use the point-splitting technique to derive the
chiral anomalies. In Section \ref{pilfavc}, we discuss the
physical implication of the new light-front axial-vector current.
A conclusion is given in section \ref{sum}.

\section{Chiral transformation in the light-front QCD} \label{ctilfqcd}


Before we discuss chiral symmetry in light-front QCD, we consider first
a free fermion theory for an illustration. Explicitly, the free fermion
Lagrangian in the light-front coordinates with $x^+=x^0+x^3$ as the "time"
direction can be written as
\be
{\cal L} = \bar{\psi}(i\not\!\partial-m)\psi
         = -\psi_+^{\dagger}\frac{\Box^2+m^2}{i\partial^+}\psi_+,
\label{FFL}
\ee
where $\psi_+$ is called the up component of fermion field:
$\psi = \psi_+ + \psi_-$ with the definition $\psi_{\pm} \equiv
{1\over4}\gamma^{\mp}\gamma^{\pm}\psi \equiv \Lambda_\pm \psi$
$(\gamma^{\pm} = \gamma^0\pm\gamma^3)$, and the down component is
determined by the light-front fermion constraint
\be
    \psi_- = \frac{1}{i\partial^+}(i\vec{\alpha}_{\perp}
             \cdot\vec{\partial}_{\perp}+\beta m)\psi_+,
\label{LFC}
\ee
while $\Box^2$ is the Laplace operator:
$\Box^2 = \partial^+\partial^- -\partial_{\perp}^2$, and the
operator $\frac{1}{\partial^+}$ is defined to satisfy the identity
$(\frac{1}{\partial^+})\partial^+ = \partial^+\frac{1}{\partial^+} = 1$.
An explicit regularization of this operator is given by \cite{ZH}
\be \label{part+}
     \left(\frac{1}{\partial^+}\right)f(x^+,x^-,\vec{x}_{\perp})
  =  \frac{1}{4}\int^{\infty}_{-\infty}\d x'^-
     \epsilon(x^--x'^-)f(x^+,x'^-,\vec{x}_{\perp}),
\ee
and $\epsilon(x) = 1,0,-1$ for $x > 0, = 0, < 0$.

Eq.~(\ref{FFL}) is obvious chiral invariant regardless of that the
fermion is massive or massless. This is because the light-front Lagrangian
in eq.~(\ref{FFL}) does not contain explicitly
gamma matrix and therefore is invariant under chiral transformations.
This contradicts with eq.~(\ref{DIVEQ}) that the axial-vector current
is not conserved. Moreover, the chiral charge in light-front $Q_5^L$
is defined by
\be
    Q_5^L \equiv \frac{1}{2}\int\d x^-\d^2x_{\perp}j_5^+
          =      \int\d x^-\d^2x_{\perp}\psi_+^{\dagger}\gamma_5\psi_+
\label{CC}
\ee
which commutes with the free-fermion light-front Hamiltonian
\be
    \partial^-Q_5^L = [Q^L_5,H_{LF}] = 0,
\label{CCH}
\ee
where
\be
    H_{LF} &=& \int\d x^-\d^2x_{\perp}(i\psi_+^{\dagger}\partial^-
               \psi_+-{\cal L})
            =  \int\d x^-\d^2x_{\perp}\Big(\psi_+^{\dagger}
               \frac{-\nabla_{\perp}^2+m^2}{i\partial^+}\psi_+\Big).
\ee
In other words, the light-front Hamiltonian of a free massive
fermion theory is chiral symmetric. Thus, chiral symmetry in
light-front field theory is apparently inconsistent that the free
massive light-front Lagrangian is chiral invariant and the chiral
charge is conserved but the axial-vector current $j_5^\mu$ is not
[see eq.~(\ref{DIVEQ})].

In ref. \cite{WWHZPG}, it has been argued that this inconsistency
of the conserved light-front chiral charge with the nonconserved
axial-vector current occurs due to an improper treatment of the
zero longitudinal momentum (i.e., so-called zero modes) in
light-front theory. From the current divergence equation, the
light-front time derivative of the chiral charge is proportional
to $\int\d x^-\d^2x_\perp\bar\psi\gamma_5\psi$. When the fields in
this integral are expanded in terms of momentum eigenstates, the
diagonal terms $b^\dagger b$ and $d^\dagger d$, where $b$ and $d$
are the quark and antiquark annihilation operators, vanish;
namely, the matrix elements multiplying them vanish. Moreover, the
off-diagonal terms vanish if the zero modes are absent. Thus, it
seems to be the absence of zero modes that makes it possible for
the light-front chiral charge to be conserved irrespective of the
nonconservation of the axial-vector current for free massive
fermions \cite{WWHZPG}. Since the longitudinal momentum of any
particle cannot be negative in light-front theory, the light-front
vacuum is only occupied by particles with zero longitudinal
momentum. Undoubtedly, light-front longitudinal zero modes play an
important role in understanding nontrivial vacuum structures of
field theory. However, the above inconsistency occurs in a free
theory. The vacuum of a free theory is trivial no matter whether
there exist zero modes or not, and the explicit chiral symmetry
breaking of a free massive fermionic theory must be irrelevant to
zero modes.

In fact, the apparent inconsistency between eq.~(\ref{DIVEQ}) and
eq.~(\ref{CCH}) originates from the improper definition of the
axial-vector current from the chiral transformation in light-front
theory \cite{DM}. According to Noether's theorem, a current is
defined from the associated continuous transformation. In
light-front theory, as we have seen the fermion field $\psi$ is
divided into $\psi_+$ and $\psi_-$, only one of them is a
dynamical independent variable. It is easy to check that the usual
chiral transformation,
\be \label{CT}
    \psi(x) \mapsto \psi'(x) = {\rm e}^{-i\theta\gamma_5}\psi(x)
\ee
is inconsistent with eq.~(\ref{LFC}). The light-front fermion
constraint eq.~(\ref{LFC}) is derived from the fermion equation of
motion and therefore it must be obeyed. Thus, the chiral
transformation can only be applied to $\psi_+$,
\be
    \psi_+(x) \mapsto \psi'_+(x) = {\rm e}^{-i\theta\gamma_5}\psi_+(x)
\label{DCT}
\ee
and the chiral transformation for $\psi_-$ is determined by
eq.~(\ref{LFC}),
\be
\psi_-(x) \mapsto \psi'_-(x)
          & = &   \frac{1}{i\partial^+}(i\vec{\alpha}_{\perp}
                  \cdot\vec{\partial}_{\perp}+\beta m)\psi'_+(x) \nn\\
          & = &   {\rm e}^{-i\theta\gamma_5}\psi_-
                  +m\gamma^0\left({\rm e}^{-i\theta\gamma_5}
                  -{\rm e}^{i\theta\gamma_5}\right)\frac{1}{i\partial^+}\psi_+(x).
\label{DCT-}
\ee
To obtain a consistent light-front axial-vector current from
the above chiral transformation, we start from the light-front
Lagrangian eq.~(\ref{FFL}) for an infinitesimal transformation i.e.,
$\psi'_+ = (1-i\theta \gamma_5)\psi_+$. The result is
\be \label{ctl}
{\cal L}'(x)={\cal L}(x)-\theta \partial_{\mu}\tilde{j}_5^{\mu},
\ee
where $\tilde{j}_5^{\mu}$ is given by
\be
\tilde{j}_5^{\mu} & = & \bar{\psi}\gamma^{\mu}\gamma_5\psi
                        -m\bar{\psi}\gamma^{\mu}\gamma_5\gamma^+
                        \frac{1}{i\partial^+}\psi \nn\\
                  & = & j_5^{\mu}-\hat j_5^\mu. \label{AVC}
\ee
This is the new axial-vector current with respect to the
light-front chiral transformation eq.~(\ref{DCT}). Eq.~(\ref{ctl}) shows
that under the light-front chiral transformation, the free fermion Lagrangian
is invariant up to a total derivative term. By the Noether's theorem, we have
\be \label{CAVC}
    \partial_{\mu} \tilde{j}_5^{\mu} = 0.
\ee
In other word, for a free fermion theory in light-front, the
chiral symmetry is conserved and the conserved axial-vector
current is given by $\tilde{j}_5^{\mu}$ of eq.~(\ref{AVC}).

Furthermore, it is remarkable to see that, although the
light-front axial-vector current $\tilde{j}_5^\mu$ is different
from the conventional one $j_5^\mu$, the $\tilde{j}_5^+$ component
is the same as $j_5^+$ component, and so does the light-front
chiral charge,
\be
    \tilde{Q}_5^L \equiv \frac{1}{2} \int \d x^- \d^2x_{\perp} \tilde{j}_5^+
                  =      \int \d x^- \d^2x_{\perp} j_5^+ = Q_5^L .
\label{CC1}
\ee
This is why even if we define the light-front
chiral charge $Q^L_5$ from the conventional definition in the
beginning, it is still a conserved one. The conservation of chiral
charge in light-front theory is just a consequence of
eq.~(\ref{CAVC}). Physically, this result is easy to be
understood. A free light-front fermion theory is chiral invariant
since the light-front chiral charge operator is identical to the
fermion's helicity operator, no matter the fermion is massive or
massless. Explicitly, the chirality of a fermion is characterized
by the chiral charge in light-front theory, $\tilde{Q}_5^L$ defined
by eq.~(\ref{CC1}). In the momentum space,
\be\label{CCP}
\tilde{Q}_5^L &=& \int\frac{\d p^+\d^2\vec{p}_{\perp}}{2(2\pi)^3}
                  \sum_{\lambda=\pm1/2}2\lambda
                  \left[b^{\dagger}(p,\lambda)b(p,\lambda)
                  +d^{\dagger}(p,-\lambda)d(p,-\lambda)\right],
\ee
where $\lambda$ denotes the helicity. Thus, the chirality is identical to
the helicity in light-front theory. If the Lagrangian does not contain
helicity flip interactions, the helicity as well as the chirality
are always conserved.

Moreover, from eqs.~(\ref{AVC}) and (\ref{CAVC}), one can easily
reproduce eq.~(\ref{DIVEQ}) i.e.,
\be
    \partial_{\mu}j_5^{\mu} = 2im\bar{\psi}\gamma_5\psi. \nn
\ee
Thus we have reached to a consistent solution that a free fermion theory
in light-front, no matter if it is a massless or massive theory, is chiral
invariant but the light-front axial-vector current must be defined in terms
of eq.~(\ref{AVC}). An alternative discussion of the light-front axial-vector
current based on the definition of Noether current for a free theory has also
been given by Mustaki \cite{DM}. In recent years, one has also use such new
light-front axial-vector to study chiral dynamics in chiral Yukawa model
and Nambu-Jona-Lasinio model \cite{IM, MV2001}. However, up to date,
none has explored carefully about the chiral symmetry for light-front QCD
from this point of view.


Now we consider QCD. We start from light-front QCD Lagrangian
(namely, the QCD Lagrangian expressed in terms of all the physical
light-front field variables \cite{ZH})
\be
{\cal L}_{\rm LFQCD} = \bar{\psi}(i\not\!\!{D}-m)\psi
                       -\frac{1}{2}{\rm Tr}(F_{\mu\nu}F^{\mu\nu}).
\ee
where we shall take the light-front gauge $A^+_a=0$, then $\psi = \psi_+ + \psi_-$
such that $\psi_-$ is determined from the QCD Dirac equation,
\be
\psi_- = \frac{1}{i\partial^+}(i\vec{\alpha}_{\perp}\cdot\vec{\partial}_{\perp}
         +g\vec{\alpha}_{\perp}\cdot\vec{A}_{\perp}+\beta m)\psi_+.
\label{LFQCDFC}
\ee
In light-front QCD, the chiral transformation of $\psi_+$ is defined as the same
as eq.~(\ref{DCT}) but the corresponding transformation of $\psi_-$ is determined
from eq.~(\ref{LFQCDFC}).
The light-front axial-vector current can be obtained through an infinitesimal
chiral transformation,
\be \label{LFCT}
\psi_+(x)\mapsto\psi'_+(x)=(1-i\gamma_5\theta)\psi_+(x)
\ee
to the LFQCD action ${\cal S}=\int\d^4\tilde{x}{\cal L}_{\rm LFQCD}$.
Under these transformations, we find that
\be \label{ctqcdl}
  {\cal L}'_{\rm LFQCD}(x) &=& {\cal L}(x)-\theta\left[
                               \partial_{\mu}\tilde{j}_5^{\mu}
                               -2mg\psi_+^{\dagger}\gamma_5
                               \left(\frac{1}{\partial^+}
                               \vec{\gamma}_{\perp}\cdot\vec{A}_{\perp}
                               \right)\psi_+\right]
\ee
where the light-front axial-vector current $\tilde{j}^\mu_5$ for QCD is
given the same as eq.(\ref{AVC}),
\be \label{AVC1}
\tilde{j}_5^{\mu} & = & \bar{\psi}\gamma^{\mu}\gamma_5\psi
                        -m\bar{\psi}\gamma^{\mu}\gamma_5\gamma^+
                        \frac{1}{i\partial^+}\psi \nn\\
                  & = & j_5^{\mu}-\hat j_5^\mu,
\ee
except that now the constraint of $\psi_-$ is determined by eq.~(\ref{LFQCDFC}).
Compare with eq.~(\ref{ctl}) and eq.~(\ref{ctqcdl}), one can see that the
light-front QCD Lagrangian is not invariant under the light-front chiral
transformation eq.~(\ref{LFCT}). It is easy to check by equations of motion
that the divergence of $\tilde{j}_5^\mu$ is given by
\be
\partial_{\mu}\tilde{j}_5^{\mu} & = & -2mg\psi^{\dagger}_+\gamma_5 \left(
                                      \frac{1}{\partial^+} \vec{\gamma}_{\perp}
                                      \cdot \vec{A}_{\perp}\right)\psi_+\nn\\
                                & = & mg\bar{\psi}\gamma_5\gamma^+ \left(
                                      \frac{1}{\partial^+} \vec{\gamma}_{\perp}
                                      \cdot\vec{A}_{\perp}\right)\psi.
\label{LFDIVEQ}
\ee
The right-hand side of the above equation is only related
to the term in the light-front QCD Lagrangian that flips quark's helicity
through its interaction with gluons. It tells us that the manifestation of
chiral symmetry breaking in light-front QCD is different from that in
equal-time formalism. Here the chiral symmetry of light-front QCD is broken
due to quark-gluon interaction involving a helicity flip process.
Meanwhile, from the new light-front axial-vector current,
one can check that eq.~(\ref{LFDIVEQ}) is still consistent with the
familiar divergent equation of the conventional axial-vector current
eq.~(\ref{DIVEQ}), namely,
\be
\partial_\mu \tilde{j}_5^\mu & = & \partial_\mu j_5^\mu - m\partial_\mu
                                   ( \bar{\psi}\gamma^{\mu}\gamma_5\gamma^+
                                   \frac{1}{i\partial^+}\psi ) \nn\\
                             & = & mg\bar{\psi}\gamma_5\gamma^+ \left(
                                   \frac{1}{\partial^+}\vec{\gamma}_{\perp}
                                   \cdot\vec{A}_{\perp}\right)\psi.
\ee
which leads to the familiar form in equal-time formalism.
\be
    \partial_{\mu} j_5^{\mu} = 2 i m \bar{\psi} \gamma_5 \psi . \nn
\ee
Furthermore, we can see from eq.~(\ref{AVC1}) that the
light-front axial-vector current is just the conventional
axial-vector current subtracting an additional term $\hat{j}^\mu_5$.
As we will discuss in Sec. IV, the operator $\hat{j}_5^{\mu}$ is related
to the pion pole contribution in hadronic transition processes.

To further check the consistency of the light-front chiral
transformation obtained here, we may calculate the commutator
of $\psi_+$ and $\tilde{Q}^L_5=Q^L_5$ in light-front theory:
using the light-front canonical quantization rule,
\be
    \{ \psi_+(x) , \psi_+^\dagger(y) \}_{x^+=y^+} = \Lambda_+ \delta^3(x-y),
\label{LFFC}
\ee
it is easy to find that
\be
[\psi_+(x),Q_5^L] = \gamma_5 \psi_+(x) .  \label{LFCCC}
\ee
This corresponds to the infinitesimal chiral transformation of eq.~(\ref{DCT}).
The corresponding commutator of $\psi_-$ can be evaluated using
eqs.~(\ref{LFQCDFC}), (\ref{LFFC}) and (\ref{LFCCC}), with the result
\be
    [\psi_-(x),Q_5^L]= \gamma_5 \psi_-(x) - m \gamma_5\gamma^+
                       {1 \over i\partial^+} \psi_+(x)
\ee
which gives the infinitesimal chiral transformation of eq.~(\ref{DCT-}).
Then the transformation of quark field under $Q^L_5$ is given by
\be
    [\psi(x),Q_5^L] = \gamma_5 \psi(x) - m \gamma_5\gamma^+
                      {1 \over i\partial^+} \psi_+(x) .
\ee
Note that $\tilde{Q}^L_5 = Q^L_5$, these results
are consistent with the definition of the light-front chiral
transformation, eqs.~(\ref{DCT} - \ref{DCT-}). It shows that the usual
chiral transformation eq.~(\ref{CT}) (which corresponds to
$[\psi(x),Q_5] = \gamma_5 \psi(x)$) is invalid for light-front theory.

\section{Chiral anomalies} \label{ca}

Before we discuss the physical implication of the new obtained light-front
axial-vector current for QCD, we should derive in this section the anomaly
associated with this light-front axial-vector current.
In equal-time frame, the chiral anomaly is originated from vacuum effect.
In the light-front theory, the vacuum is occupied only by $k^+=0$ particles.
With the help of infrared regularization of longitudinal light-front momentum,
the light-front vacuum is ensure to be trivial as the bare vacuum \cite{WWHZPG,ZH}.
Now we use the different vacuum structure and a new light-front axial-vector current
to derive the chiral anomaly.
There are three methods to derive the chiral anomaly of axial-vector current,
namely the point-splitting technique \cite{CRH,RJKJ,TJZW},
the triangle diagrams \cite{TJZW,SLA,AB,ABO} and the path-integral method \cite{KF}.
The point-splitting technique has been used to calculate chiral anomaly in
two and four dimensional light-front QED, where the conventional axial-vector
current is used \cite{MV2K}. In our work, we shall use the point-splitting technique
to calculate chiral anomaly of the associated new light-front axial-vector
current in light-front QCD.

The light-front axial-vector current is also a composite operator built out
of quark fields. The products of local operators are often singular.
To regularize the axial-vector current, we introduce a small separation of
the two quarks and insert the Wilson line to keep the operator being locally
gauge invariant. At the end of calculation, we then carefully take $\epsilon$
to zero. It must be pointed out that after we choose the gauge $A_a^+ = 0$,
the gauge degrees of freedom are fixed except for possible residual gauge
transformations. The requirement of gauge invariance means that of the residual
gauge transformation,
\be
 && \psi'(x)  =  {\rm e}^{i\alpha(x)T^a}\psi(x), \nn\\
 && A'^\mu_a(x)  =  A_a^\mu(x)+\frac{1}{g}\partial^\mu\alpha_a(x)
                    +f^{abc}A_b^\mu(x)\alpha_c(x), \nn
\ee
where $\alpha_a(x)$ is the gauge function that obeys $\partial^+\alpha_a(x)  =  0$
such that the light-front gauge $A_a^+ = 0$ is guaranteed.
Thus, from eq.~(\ref{AVC1}), the gauge invariant light-front axial-vector
current is given by
\be \label{PCAC1}
    \tilde{j}_5^{\mu} &=& \lim_{\epsilon\to0}\Big\{\bar{\psi}(x+\epsilon/2)
                          \gamma^{\mu}\gamma_5 {\cal W}(A) \psi(x-\epsilon/2) \nn\\
                      && -m\bar{\psi}(x+\epsilon/2)\gamma^{\mu}\gamma_5\gamma^+
                          {\cal W}(A) \frac{1}{i\partial^+}\psi(x-\epsilon/2)\Big\},\nn\\
\ee
where
$
    {\cal W}(A) = \exp\Big[ig\int_{x-\epsilon/2}^{x+\epsilon/2}\d z\cdot A(z)\Big]
$
is the Wilson line. We write down the each component of above current in terms of
$\psi_+$ and $\psi_-$
\be
\tilde{j}_5^{+}  & = & 2\lim_{\epsilon\to0}\Big\{\psi_+^\dagger (y_+)
                       \gamma_5 {\cal W}(A) \psi_+(y_-)\Big\}, \nn\\
\tilde{j}_5^{-}  & = & 2\lim_{\epsilon \to 0}\Big\{\psi_-^\dagger (y_+)
                       \gamma_5 {\cal W}(A) \psi_-(y_-)
                       - 4 m \psi_-^\dagger (y_+)\gamma_5\gamma^0
                       {\cal W}(A) \frac{1}{i\partial^+}\psi_+(y_-)\Big\},\nn\\
\tilde{j}_5^{~i} & = & \lim_{\epsilon \to 0}\Big\{\psi_+^\dagger(y_+)
                       \gamma^0\gamma^i\gamma_5 {\cal W}(A) \psi_-(y_-)
                       +\psi_-^\dagger(y_+)
                       \gamma^0\gamma^i\gamma_5 {\cal W}(A) \psi_+(y_-) \nn\\
                 &   & - 2 m \psi_+^\dagger(y_+)\gamma^i\gamma_5
                       {\cal W}(A) \frac{1}{i\partial^+}\psi_+(y_-)\Big\},
\ee
where $y_\pm=x\pm\epsilon/2$ and $\psi_-$ is given by eq.~(\ref{LFQCDFC}).
Then we calculate directly the divergence of the above light-front axial-vector
current,
$
\partial_{\mu}\tilde{j}_5^{\mu} = \frac{1}{2}\partial^-\tilde{j}_5^+
                                  +\frac{1}{2}\partial^+\tilde{j}_5^-
                                  -\partial^i\tilde{j}_5^i. \nn
$
Note that these derivative terms, $\partial^-\psi_+$, $\partial^-\psi_-$
and their complex conjugate, can be substituted by using
the light-front fermion constraint eq.~(\ref{LFQCDFC}) and
the equation of motion
\be
(i\partial^-+gA^-)\psi_+ = (i\vec\alpha_\perp+g\vec\alpha_\perp\cdot\vec{A}_\perp
                           +\beta m)\psi_-,
\ee
where $A^-$ satisfies light-front gluon constraint:
\be \label{lfgc}
A_a^- = 2\frac{\partial^i}{\partial^+}A_a^i-2g\Big(\frac{1}{\partial^+}\Big)^2
        (f^{abc}A_b^i\partial^+A_c^i+2\psi_+^\dagger T^a \psi_+).
\ee
A tedious but straightforward calculation shows that
\be \label{psfde}
     \partial_{\mu}\tilde{j}_5^{\mu}
&=&  i g \psi_+^\dagger(y_+)\Big\{{\cal W}(A) A^-(y_-) - A^-(y_+){\cal W}(A)
     + \epsilon^\nu\partial^- A_\nu (x){\cal W}(A)\Big\}\gamma_5 \psi_+(y_-) \nn\\
&+&  i g \psi_+^\dagger(y_+)\Big\{
     {\cal W}(A) A^i(y_-) - A^i(y_+){\cal W}(A)
     + \epsilon^\nu\partial^i A_\nu (x){\cal W}(A)\Big\}
     \gamma^i\gamma^0 \gamma_5 \psi_-(y_-) \nn\\
&+&  i g \psi_-^\dagger(y_+)\Big\{{\cal W}(A)A^i(y_-)
     - A^i(y_+){\cal W}(A)
     + \epsilon^\nu\partial^i A_\nu (x){\cal W}(A)\Big\}
     \gamma^i\gamma^0 \gamma_5 \psi_+(y_-) \nn\\
&+&  i g \psi_-^\dagger (y_+)\Big\{
     \epsilon^\nu\partial^+A_\nu(y_+){\cal W}(A)\Big\}
     \gamma_5\psi_-(y_-) \nn\\
&+&  2mg \psi_+^\dagger(y_+)\gamma^i\gamma_5{\cal W}(A)
     \Big(\frac{1}{\partial^+}A^i(y_-)\Big)\psi_+(y_-) \nn\\
&+&  2mg \psi_+^\dagger(y_+) ({\cal W}(A)A^i(y_-)-A^i(y_+){\cal W}(A))
     \gamma^i\gamma_5\frac{1}{\partial^+}\psi_+(y_-) \nn\\
&+&  2mg \psi_-^\dagger(y_+)\gamma^0\gamma_5
     ( \epsilon^\nu\partial^+ A_\nu(x)){\cal W}(A)
     \frac{1}{\partial^+}\psi_+(y_-) \nn\\
&+&  2mg \psi_+^\dagger(y_+)\gamma^i\gamma_5
     ( \epsilon^\nu\partial^i A_\nu(x)){\cal W}(A)
     \frac{1}{\partial^+}\psi_+(y_-) .
\ee
Here $\epsilon^\mu$ is small, we take an expansion of the independent gauge fields
$A^i$ with
respect to $\epsilon$,
\be
      A^i(x\pm\frac{\epsilon}{2})\approx A^i(x)\pm
      \frac{\epsilon^\beta}{2}\partial_\beta A^i(x). \label{eaf}
\ee
The expansion of the dependent component of the gauge fields $A_a^-$ can be determined
from the light-front gluon constraint eq.~(\ref{lfgc}):
\be
           A_a^-(x\pm\frac{\epsilon}{2})
      &=&  2\frac{\partial^i}{\partial^+} A_a^i(x\pm\frac{\epsilon}{2}) \nn\\
      & &     - 2g\Big(\frac{1}{\partial^+}\Big)^2 [f^{abc}A_b^i(x\pm\frac{\epsilon}{2})
           \partial^+A_c^i(x\pm\frac{\epsilon}{2})+2\psi_+^\dagger(x\pm\frac{\epsilon}{2})
           T^a \psi_+(x\pm\frac{\epsilon}{2})] \nn\\
&\approx&  2\frac{\partial^i}{\partial^+} A_a^i(x) - 2g\Big(\frac{1}{\partial^+}\Big)^2
           [f^{abc}A_b^i(x)\partial^+A_c^i(x) +2\psi_+^\dagger(x) T^a \psi_+(x)] \nn \\
      & &  \pm \frac{\epsilon^\beta}{2}\partial_\beta\Big\{2\frac{\partial^i}{\partial^+}
            A_a^i(x) - 2g\Big(\frac{1}{\partial^+}\Big)^2
           [f^{abc} A_b^i(x)\partial^+A_c^i(x)+2\psi_+^\dagger(x) T^a \psi_+(x)]\Big\}\nn \\
      &=&  A_a^-(x)\pm \frac{\epsilon^\beta}{2}\partial_\beta A_a^-(x).
\ee
Thus, keeping in mind the expansion is taken up to the order $\epsilon$:
\be
    & &  A_\alpha(x\pm\frac{\epsilon}{2})\approx A_\alpha(x)\pm
        \frac{\epsilon^\beta}{2}\partial_\beta A_\alpha ,\nn\\
    & & {\cal W}(A)\approx 1+ig\epsilon^\alpha A_\alpha (x),
\ee
we can reduce eq.~(\ref{psfde}) to
\be
                  \partial_{\mu}\tilde{j}_5^{\mu}
             &=&  \lim_{\epsilon\to0} \Big\{
                  ig\epsilon^{\nu}F_{a\mu\nu}\bar\psi(x+\frac{\epsilon}{2})
                  \gamma^{\mu}\gamma_5T^a\psi(x-\frac{\epsilon}{2})\nn \\
             & &  -mg\epsilon^{\nu}F_{a~~\nu}^{~i}\bar\psi(x+\frac{\epsilon}{2})
                  \gamma^+\gamma^{i}\gamma_5T^a\psi(x-\frac{\epsilon}{2})\nn \\
             & &  -mg\epsilon^{\nu}F_{a~~\nu}^{~+}\bar\psi(x+\frac{\epsilon}{2})
                  \gamma^-\gamma_5\gamma^0T^a\frac{1}{\partial^+}
                  \psi(x-\frac{\epsilon}{2})\Big\}\nn \\
             & &  +\lim_{\epsilon\to0} \Big\{2mg\psi_+^{\dagger}(x+\frac{\epsilon}{2})
                  \gamma^i\gamma_5 {\cal W}(A)\Big(\frac{1}{\partial^+}
                  A^i(x-\frac{\epsilon}{2})\Big)\psi_+(x-\epsilon/2) \Big\},
\label{PAVC}
\ee
where $T^a$ is a color SU(3) generator and
$F_{\mu\nu}=\partial_\mu A_{\nu}-\partial_\nu A_{\mu}-ig [A_\mu,A_\nu]$.
The last term in eq.~(\ref{PAVC}) is regular as well as $\epsilon$ approaching to zero,
which gives us the divergence in eq.~(\ref{LFDIVEQ}).
Consider the vacuum expectation value of $\partial_{\mu}\tilde{j}_5^{\mu}$.
The first term of eq.~(\ref{PAVC}) is
\be
&&  ig\epsilon^{\nu}F_{a\mu\nu}\langle0|\bar\psi(x+\frac{\epsilon}{2})
    \gamma^{\mu}\gamma_5T^a\psi(x-\frac{\epsilon}{2})|0\rangle \nn\\
&&  =ig\epsilon^{\nu}F_{a\mu\nu}{\rm Tr}\Big\{ T^a{\rm Tr}\Big\{\gamma^{\mu}\gamma_5
    \tilde{G}(x-\frac{\epsilon}{2},x+\frac{\epsilon}{2})\Big\}\Big\}.
\label{VEV}
\ee
With the existence of background field, the propagator of fermions can
be expanded as
\be
    \tilde{G}(x-\frac{\epsilon}{2},x+\frac{\epsilon}{2})
&=& \langle0|\psi(x-\frac{\epsilon}{2})\bar\psi(x+\frac{\epsilon}{2})|0\rangle \nn\\
&=& \tilde{S}(-\epsilon)+\int\d^4y\tilde{S}(x-\frac{\epsilon}{2}-y)(ig\not\!\!A(y))
    \tilde{S}(y-x-\frac{\epsilon}{2}),
\ee
where
\be
       \tilde{S}(x) = \int\frac{\d^4p}{(2\pi)^4}{\rm e}^{-ip\cdot x}\left[
                      \frac{i(\not\!p+m)}{p^2-m^2+i\varepsilon}
                      -\frac{1}{2}i\frac{\gamma^+}{p^+}\right]
\ee
is the light-front propagator of quark field \cite{CY}. Thus, eq.~(\ref{VEV}) becomes
\be
& &  \langle0| \psi(x+\frac{\epsilon}{2})\gamma^{\mu}
     \gamma_5T^a\psi(x-\frac{\epsilon}{2})|0\rangle \nn\\
& &  = -{\rm Tr}\{T^aT^b\}\int\frac{\d^4p}{(2\pi)^4}
     \int\frac{\d^4k}{(2\pi)^4}e^{-ik\cdot\epsilon}e^{-ip\cdot(x+\epsilon/2)}
     {\rm Tr}\Big\{\gamma^{\mu}\gamma_5\tilde{S}(p+k)(ig\not\!\!A^b(p))
     \tilde{S}(k)\Big\} \nn \\
& &  = ig{\rm Tr}\{T^aT^b\}\int\frac{\d^4p}{(2\pi)^4}
     \int\frac{\d^4k}{(2\pi)^4}e^{-ik\cdot\epsilon}e^{-ip\cdot(x+\epsilon/2)}\nn\\
& &  \Big\{4i\epsilon^{\mu\alpha\beta\rho}\frac{(p+k)_\alpha A^b_\beta(p)k_\rho}
     {[(p+k)^2-m^2+i\varepsilon][k^2-m^2+i\varepsilon]}
     -2i\epsilon^{\mu\alpha\beta+}\frac{(p+k)_\alpha A^b_\beta(p)}
     {[(p+k)^2-m^2+i\varepsilon]k^+}\nn\\
& &  +2i\epsilon^{\mu\alpha\beta+}\frac{A^b_\beta(p)k_\alpha}
     {[k^2-m^2+i\varepsilon]k^+}\Big\}.
\label{AN}
\ee
As $\epsilon \to 0$, the three terms in eq.~(\ref{AN}) are proportional to
$k$, $k^2$ and $k^2$ respectively, in the integration of k. Thus, the main
contribution comes from the large $k$ region, namely $p<<k$. With such a
consideration \cite{PS}, the last two terms of eq.~(\ref{AN}) cancel each
other and the first term is reduced to
\be
    \int\frac{\d^4p}{(2\pi)^4}e^{-ip\cdot x} p_\alpha A^b_\beta(p)
    \int\frac{\d^4k}{(2\pi)^4}e^{-ik\cdot\epsilon}
    \frac{k_\rho}{k^4}
  = \partial_\alpha A^a_\beta(x)
    \Big(\frac{-i}{8\pi^2}\frac{\epsilon_\rho}{\epsilon^2}\Big).
\ee
One can easily verify that the above result is singular as $\epsilon \to 0$,
as we mention in the front of this section.
But the combination with the term $\epsilon^\nu$ in eq.~(\ref{VEV}) gives
a finite contribution:
\be
      \lim_{\epsilon\to 0}\frac{\epsilon^\nu\epsilon_\rho}{\epsilon^2}
    = \frac{1}{4} g^\nu_{~~\rho}.
\ee
Substituting these results into eq.~(\ref{AN}), we have
\be
       ig\epsilon^{\nu}F_{a\mu\nu}\langle0|\psi(x+\frac{\epsilon}{2})
       \gamma^{\mu}\gamma_5T^a\psi(x-\frac{\epsilon}{2})|0\rangle
    =  -\frac{g^2}{16\pi^2}\epsilon^{\alpha\beta\mu\nu}{\rm Tr}
       \{F_{\alpha\beta}F_{\mu\nu}\}.
\ee
A similar calculation shows that the other two terms in eq.~(\ref{PAVC})
give no contribution. Thus, the anomalous term for the light-front
axial-vector current is identical to that in equal-time formalism.
Then the divergence of light-front axial-vector current can be written
as follows:
\be
      \partial_{\mu}\tilde{j}_5^{\mu} = mg\bar{\psi} \gamma_5 \gamma^+
      \left(\frac{1}{\partial^+}\vec{\gamma}_{\perp}\cdot\vec{A}_{\perp}\right)\psi
      -\frac{g^2}{16\pi^2}\epsilon^{\alpha\beta\mu\nu}{\rm Tr}
      \{F_{\alpha\beta}F_{\mu\nu}\}.
\label{DAVC}
\ee
Using eq.~(\ref{AVC}) and eq.~(\ref{DAVC}) can be reduced to the
conventional form:
\be
    \partial_{\mu}j_5^{\mu}=2im\bar{\psi}\gamma_5\psi-\frac{g^2}{16\pi^2}
    \epsilon^{\alpha\beta\mu\nu}{\rm Tr}\{F_{\alpha\beta}F_{\mu\nu}\},
\label{DAVC2}
\ee
namely, eq.~(\ref{DAVC}) and eq.~(\ref{DAVC2}) are again consistent each other
even if the chiral anomaly is included.

\section{Physical implication of light-front axial-vector current}\label{pilfavc}

Now we shall discuss the physical implication of the new light-front
axial-vector current. In this section, we study physical implication in
light-front QCD from two aspects, the spontaneous symmetry breaking (SSB)
and the hadronic transition matrix element of axial-vector current.
In equal-time formulation,
the divergence of axial-vector current eq.~(\ref{DIVEQ}) is
considered as an interpolating pion field through the partially
conserved axial-vector current (PCAC) hypothesis:
\be
    \partial_\mu j_5^\mu(x) = m_\pi^2f_\pi\phi(x),
\label{PCAC}
\ee
where $\phi(x)$ is the pion field operator, $f_\pi$ the pion decay
constant and $m_\pi$ the pion mass. Comparing eq.~(\ref{DIVEQ})
with eq.~(\ref{PCAC}), one can see that the composite operator
of quark fields $\bar\psi i\gamma_5\psi$ is the interpolating
pion field, and the limit of quark mass $m\to 0$ implies
$m_\pi=0$ i.e., $\pi$ is massless in the massless quark
theory, as a result of the spontaneous chiral symmetry
breaking. This is also manifested from the commutation of
the static chiral charge $Q_5$ [defined in equal-time formalism
$Q_5=\int\d^3x j_5^0(x)$] with the interpolating pion field operator:
\be
    [ Q_5 , \bar\psi(x) i \gamma_5 \psi(x) ]_{x^0=0} = -2 i \bar\psi(x) \psi(x).
    \label{condensate}
\ee
The nonzero value of the vacuum expectation
$\langle\bar\psi\psi\rangle$ (chiral condensate) is the character
of spontaneous chiral symmetry breaking
($Q_5|\Omega\rangle=|\Phi\rangle$), namely the chiral charge
creates massless Goldstone bosons from the vacuum
as a consequence of spontaneous chiral symmetry breaking.
These discussion seems to be at variance with the light-front
framework because of $\tilde{Q}_5^{L}|0\rangle = 0$.
To be more explicit, we check the commutation of the light-front chiral
charge ($\tilde{Q}_5^{L}$) with the interpolating pion field
($\bar\psi i \gamma_5 \psi$). The result is
\be
           [\tilde{Q}_5^{L},\bar\psi(x) i \gamma_5\psi(x)]_{x^+=0}
      &=&  -2i\bar\psi(x)\psi(x) \nn\\
      & &  + m \left[\bar\psi(x) \gamma^+
           \left(\frac{1}{\partial^+}\psi(x)\right)
          -\left(\frac{1}{\partial^+}\bar\psi(x) \right)
          \gamma^+\psi(x)\right].
\label{LFcondensate}
\ee
Comparing with eq.~(\ref{condensate}), there is an extra term
$m \left[\bar\psi(x) \gamma^+\left(\frac{1}{\partial^+}\psi(x)\right)
-\left(\frac{1}{\partial^+}\bar\psi(x) \right)\gamma^+\psi(x)\right]$.
The vacuum expectation value of the left-hand side of eq.~(\ref{LFcondensate})
must vanish due to the fact that the light-front chiral charge annihilates
the vacuum, $\tilde{Q}_5^{L}|0\rangle = 0$. Thus the right-hand side must vanishes
as well. This leads to
\be \label{nzc}
              \langle 0|\bar\psi\psi|0\rangle
            = \frac{-im}{2} \langle 0|\left[\bar\psi\gamma^+
              \left(\frac{1}{\partial^+}\psi\right)
              -\left(\frac{1}{\partial^+}\bar\psi\right)
              \gamma^+\psi\right]|0\rangle ,
\ee
namely the quark condensate in light-front theory can be non-zero even if the chiral
charge annihilates the vacuum. This is no surprise because the right-hand side of
eq.~(\ref{LFcondensate}) contains $(1/\partial^+)$ and therefore is infrared singular.
The renormalization of the infrared divergence (related to the light-front zero mode
in QCD) can give $\langle 0|\bar\psi\psi|0 \rangle$ a nonzero value even in the chiral
limit $m \to 0$.
However, for Abelian QED, since $\psi$
(electron) is a physical particle field, it should vanish in spacial
infinity. Thus, eq.~(\ref{part+}) for $1/\partial^+$ which leads to an
antisymmetric boundary condition for $\psi$ at the longitudinal
infinity cannot apply to QED, and as a result, the light-front
longitudinal infrared problem in QED is trivial (within the
perturbation theory) and cannot generate a non-zero chiral
condensate of eq.~(\ref{nzc}).

This result is analogous to Higgs mechanism of SSB in the light-front formulation of
the standard model \cite{SBPRD} that moving the complicity of the vacuum state into
the operator level. However, the straightforward Higgs mechanism of SSB by a simple shift
of the scalar field cannot be applied to QCD. Here, such a realization is naturally
carried out from the new light-front axial-vector current derived from the consistent
light-front chiral transformation. While the possible non-vanishing quark condensate is
associated with the infrared divergence in light-front theory, namely the renormalization
treatment of the light-front zero mode in QCD. Unfortunately, a practical scheme of the
light-front infrared renormalization  for QCD still remains to be solved \cite{Wilson}.


On the other hand, an arbitrary hadronic transition matrix
element of axial-vector
current can be separated into a pole term plus a non-pole term:
\be \label{meac}
    \langle B | j^\mu_5 (0) |A \rangle
 =  {if_\pi q^\mu \over q^2 - m^2_\pi}
    \langle B, \pi |A\rangle
    + \langle B | j^\mu_5 |A \rangle_N ,
\ee
where $p_{A(B)}$ is the momentum carried by hadron A(B) and
$q = p_A - p_B$.
This indicates that the axial-vector current
$j^\mu_5$ contains two resources, one describes the pion mode in
vacuum, and the other describes hadronic transitions involving
pion. As it has been discussed in \cite{CHKT},
using the PCAC hypothesis, eq.~(\ref{PCAC}), one can show from
eq.~(\ref{meac}) that
\be
    \langle B \pi | A \rangle = iq_\mu  \langle B | j^\mu_5
                                |A \rangle_N /f_\pi .
    \label{pp}
\ee
Substituting eq.~(\ref{pp}) into eq.~(\ref{meac}), we obtain
\be
    \langle B | j^\mu_5 (0) |A \rangle
 =  \Big(g^\mu_{~\nu}-{q^\mu q_\nu \over q^2 - m^2_\pi}\Big)
    \langle B | j^\nu_5 |A \rangle_N .
\label{meac1}
\ee
If we take the limit, $q^+,\vec q_\perp \to 0$, then
\be \label{mecc1}
    \langle B |Q_5^L | A \rangle
 =  \frac{\langle p_B|| p_A \rangle}{p_A^++p_B^+}\langle B | j_5^+
    |A \rangle_N \Big|_{q^+=\vec q_\perp=0} ,
\ee
where $\langle p_B|| p_A \rangle = 2(2\pi)^3p_A^+\delta(p_A^+ - p_B^+)
\delta^2(\vec p_{A\perp}- \vec p_{B\perp})$. The eq.~(\ref{mecc1}) implies that
hadronic transition matrix elements of light-front chiral charge is
only contributed by the non-pole term. In other words, the light-front
chiral charge does not contain the pole contribution.

This property has also recently been discussed phenomenologically
in the light-front formalism, namely, the axial current $j^\mu_5$
contains an interpolating field of the $\pi$-field plus a non-pole
term: $j^\mu_5 = -f_\pi \partial_\mu \pi + \tilde{j}^\mu_5$ and the
light-front chiral charge $Q^L_5 = {1\over 2}\int dx^-d^2x_\perp
\bar{j}^+_5$ is constructed only from the non-pole term
\cite{KTK}. Now, we shall show that the light-front axial-vector
current $\tilde{j}^\mu_5$ derived in this paper just corresponds
to the non-pole contribution and the interpolating $\pi$-field
($\partial^\mu \pi$) is related to
$\hat{j}^\mu_5 = j^\mu_5 - \tilde{j}^\mu_5 =
m\bar{\psi}\gamma^\mu\gamma_5\gamma^+ {1\over i\partial^+}\psi$ in
eq.~(\ref{AVC1}).

Existence of the pole term means that the chiral charge can
create a pion from vacuum or annihilate a pion [see eq.~(\ref{condensate})].
To extract the pole contribution from the axial-vector current
in light-front theory, we consider the following matrix element:
\be \label{meoavc}
                \langle 0| j_5^\mu |\pi(p) \rangle
              = \langle 0| \tilde j_5^\mu |\pi(p) \rangle
              + \langle 0| \hat j_5^\mu |\pi(p) \rangle,
\ee
here we have used the relation between the conventional and the
light-front axial-vector currents that has been shown in eq.~(\ref{AVC1}).
From eq.~(\ref{meac}), the hadronic matrix element of the conventional
axial-vector current contains both the pole term and the non-pole term.
Thus, for the matrix element $\langle 0| j_5^\mu |\pi(p) \rangle$ in
eq.~(\ref{meoavc}) only contains the pole contribution. Meanwhile,
the right-hand side of the above equation is naturally
separated into two terms based on eq.~(\ref{AVC1}). An explicit calculation
shows that the divergence of $\hat j_5^\mu$:
\be \label{dhj5}
             \langle 0| \partial_\mu \hat j_5^\mu |\pi \rangle
         &=& - mg\langle 0| \bar\psi\gamma_5\gamma^+
             \Big(\frac{1}{\partial^+}\vec\gamma_\perp\cdot\vec A_\perp\Big)\psi
             |\pi \rangle \nn\\
         & & + 2m\langle 0| \bar\psi i \gamma_5 \psi |\pi \rangle    .
\ee
The first term cancels with $\langle 0| \partial_\mu \tilde j_5^\mu |\pi \rangle$
[see eq.~(\ref{LFDIVEQ})] and the second term corresponds to the divergence
of the conventional axial-vector current. Using the PCAC hypothesis, we have
\be \label{r1}
              -ip_\mu \langle 0| \hat j_5^\mu (x)|\pi(p) \rangle
          &=& \langle 0| \partial_\mu j_5^\mu (x)|\pi(p) \rangle \nn\\
          &=&  m_\pi^2 f_\pi \langle 0| \phi(x) |\pi(p) \rangle.
\ee
Equivalently, we can write the above equation as
\be \label{hj5}
            \langle 0| \hat j_5^\mu (x)|\pi(p) \rangle
          = - f_\pi \langle 0| \partial^\mu\phi(x) |\pi(p) \rangle.
\ee
This means that the term $\hat j_5^\mu$ in eq.~(\ref{AVC1})
is related to the interpolating pion field operator, and the hadronic
transition matrix element of the light-front axial-vector current
$\tilde j_5^\mu$ is only related to the non-pole contribution.
This also explain why the light-front chiral charge annihilate
the vacuum.

\section{Summary}\label{sum}
In summary, the chiral transformation defined in light-front theory is very different
from that in equal-time frame. The light-front fermion field is divided
into up and down components, only the up component is the dynamical variable
and the down component is a constraint.
Hence, the proper chiral transformation on light-front
is defined by eq.~(\ref{DCT}). In the free fermion theory, the light-front
axial-vector current is conserved, no matter it is massless or massive.

As far we have learned from current algebra is the properties of current algebra
from free quark theory. In the early development of QCD, it was argued that the
properties remain true when the interaction between quark and gluon is switched on
because the divergence depends only on quark mass. This seems to contradict with
light-front currents which are known to be dynamical dependent. We have derived the
divergence equation of axial-vector current in light-front QCD. The light-front
axial-vector current $\tilde j_5^\mu$ is the same as the free fermion theory
except that the light-front fermion constraint is given by eq.~(\ref{LFQCDFC}).
The result presented in this paper shows that the chiral symmetry
is broken by the mixing of quark mass and quark-gluon interaction i.e. the
helicity flip interaction. We also study the quantum effect of the new
axial-vector current associated with the chiral transformation in light-front
theory. We show that the anomaly in light-front theory is consistent with the case
in equal-time frame.

Furthermore, we show in this formulism that the light-front chiral charge
annihilates the vacuum but the chiral condensate can be nonzero. Meanwhile,
the light-front axial-vector current $\tilde j_5^\mu$ can be obtained from
the axial-vector current $j_5^\mu$ by subtracting an additional term
$\hat j_5^\mu$. The term $\hat j_5^\mu$ corresponds to the pole contribution
which is very complicated in QCD, while, the light-front axial-vector current
corresponds to the non-pole contribution of the conventional axial-vector current
that smoothly describes various pion transitions even in the chiral limit.
As a conclusion, we believe that the chiral dynamics in light-front QCD can
largely simplify hadronic physics.

\acknowledgments
This work was supported in part by the National Science Council of ROC under Contract
Nos. NSC-92-2112-M-006-024 and NSC-92-2112-M-001-030.
WMZ would also like to thank A. Harindranath for the
fruitful discussion in the beginning of this work.

\end{document}